\documentclass[preprint,onecolumn,12pt,aps,prb,showpacs,showkeys]{revtex4-1}
\usepackage{graphicx}
\usepackage{color}
\usepackage{amsmath}
\usepackage{float}

\begin{document}

\preprint{RBP 03 2017 ISOT EFF}

\title{Inverse Isotope Effect in PdH(D)}

\author{S. Villa-Cort\'es and R. Baquero}

\affiliation{Physics Department, Cinvestav-IPN\\
 Av. IPN 2508 GAM, 036600 Ciudad de M\'exico, M\'exico}

\date{\today}
\begin{abstract}
In this letter, we take a new approach to the explanation of the inverse isotope effect in PdH(D). Our approach introduces two new aspects. First, we took into account the experimental evidence that at temperatures below 50 K, the crystal structure of PdH and of PdD is zincblende. Second, we studied the contribution of both, the electron-electron and the electron-phonon interactions. We used the Migdal-Eliashberg theory to perform our ab initio calculations.  We found that the electron-electron contribution is the most important one to explain the inverse isotope effect. We reproduced the experimental found values for the critical temperature and the isotope coefficient. Our analysis represents a direct and simple explanation for the inverse isotope effect in PdH(D).

\end{abstract}

%\pacs{74.62.Fj, 74.62.-c, 74.62.Yb, 74.20.-z}
%74 ....  superconductivity
% 74.62.Fj effects of pressure
% 74.62.-c transition temperature variations, phase diagrams
%74.62.yb  other effects
%74.20.-z  theories and models of superconductivity

\keywords{superconductivity, inverse isotope effect, Eliashberg theory}

\pacs{74.62.Fj, 74.62.-c, 74.62.Yb, 74.20.-z}


\maketitle

The vibration spectrum, the electron-phonon coupling and the Coulomb electron-electron repulsion determine the isotope effect. If only phonons are taken into account the Bardeen-Cooper-Schrieffer theory (BCS) \citep{PhysRev.108.1175} predicts that the transition temperature of a single- element-superconductor goes as $T_c\propto M^{-\alpha}$ where $M$ is the isotope mass and $\alpha=0.5$. If we take into account the contribution of the Coulomb electron-electron repulsion some deviations from this value can occur \citep{LEAVENS19741329}.

Quite an amount of theoretical and experimental work \citep{PhysRevB.14.3630,YUSSOUFF1995549,CHEN1989485,BROWN197599,CRESPI1992427,PhysRevLett.57.2955,PhysRevLett.35.110,PhysRevB.17.141,0022-3719-7-15-015,PhysRevB.39.4110,Kara,PhysRevB.45.12405,PhysRevB.29.4140,PhysRevLett.34.144,PhysRevB.12.117,PhysRevLett.111.177002} has been done since the discovery of the inverse isotope effect in PdH(D) $(\alpha\approx-0.3)$ without reaching to a satisfactory explanation of this phenomenon. Karakozov \textit{et al.} \citep{Kara} and Klein \textit{et al.} \citep{PhysRevB.45.12405} attribute the inverse isotope effect to anharmonic effects in the vibrational spectrum of PdH(D). Jena \textit{et al.} \citep{PhysRevB.29.4140} found that the zero-point vibration of Hydrogen and Deuterium can contribute to the isotope effect in an important way. The effect has been associated to certain electronic properties or to volume effects \citep{PhysRevLett.34.144,PhysRevB.12.117}.

Another aspect of this problem is associated to the crystal structure considered for PdH(D). Most of the work done so far considers the rocksalt crystalline structure where the hydrogen atoms are located on the octahedral sites of the fcc lattice of Palladium \citep{PhysRevB.14.3630,YUSSOUFF1995549,CHEN1989485,BROWN197599,CRESPI1992427,PhysRevLett.57.2955,PhysRevLett.35.110,PhysRevB.17.141,0022-3719-7-15-015,PhysRevB.39.4110,Kara,PhysRevB.45.12405,PhysRevB.29.4140,PhysRevLett.34.144,PhysRevB.12.117,PhysRevLett.111.177002}. Neutron diffraction techniques have been employed to study the hydrogen-atom configuration in a single-phase sample of beta-PdH at several selected temperatures. The suggested low-temperature ($T\ll55$ K) structure of this compound is one which conforms to the space group $R\bar{3}m$. This means that, depending on temperature ($T\ll55$ K), the hydrogen atoms move from their octahedral positions towards tetrahedral ones forming the zincblende structure  \citep{PhysRev.137.A483,CAPUTO-ALI,PhysRevB.78.014104,doi:10.1063/1.4901004}  For PdD something very similar occurs \citep{0295-5075-64-3-344}.  In a theoretical study, for pressures below 20 GPa at 0 K  the stable structure was found to be zincblende \citep{doi:10.1021/jp210780m}. We conclude from these observations that in the superconducting state ($T_{c}\approx9$ K) at zero pressure PdH(D) is in the zincblende and not in the rocksalt crystal structure as it has been assumed up to now. Therefore, in this work, we consider PdH(D) in the zincblende structure and will show that the inverse isotope effect can be explained in agreement with experiment starting with this hypothesis and without assuming anharmonicity.

In this letter, we show that using the Migdal-Eliashberg theory \citep{Daams1979,PhysRevB.12.905} and considering the vibrational modes to be harmonic it is possible to account for the inverse isotope effect. The importance of this work lies not only on the fact that it gives a simple and direct explanation of the inverse isotope effect but also that it can be taken as a starting point to explain the isotope effect and the change in the critical temperature of the isotope in new superconductors formed by introducing hydrogen into other elements as S, Pt, Se and P among other possibilities \citep{203,PhysRevLett.114.157004,PhysRevLett.96.017006,10.1038/srep06968,0953-2048-30-4-045011,PhysRevB.93.104526,Szczniak201730,Flores-Livas2016,doi:10.1021/acs.jpcc.5b12009,0953-2048-28-8-085018}.

BCS theory gives for a compound with several atoms the following formula for the partial isotope effect coefficients $\alpha_{i}\equiv-d\ln T_{c}/d\ln M_{i}$ where $M_{i}$ is the mass of the different atoms in the compound and $T_{c}$ the critical temperature. The total isotope effect coefficient is given by the sum of the partial ones, namely $\alpha_{tot}=\sum_{i}\alpha_{i}$. According to Migdal-Eliashberg theory, the $\alpha_{i}$ coefficients contain information on both the electron-phonon interaction and the electron-electron repulsion. If we take into account only the mass dependence of the electron-phonon interaction, Rainer and Culetto \citep{PhysRevB.19.2540} have shown that the isotope effect coefficient can be calculated from the formula

\begin{eqnarray}
\alpha_{e-ph}\left(\omega\right) & \equiv & R\left(\omega\right)\alpha^{2}F\left(\omega\right),
\end{eqnarray}
where $R\left(\omega\right)$ is given by 
\begin{eqnarray}
R\left(\omega\right) & = & \frac{d}{d\omega}\left[\frac{\omega}{2T_{c}}\frac{\delta T_{c}}{\delta\alpha^{2}F\left(\omega\right)}\right].
\end{eqnarray}
The Eliashberg function is defined as 
\begin{eqnarray}
\alpha^{2}F\left(\omega\right)= &  & \frac{1}{N\left(\epsilon_{F}\right)}\sum_{nm}\sum_{\vec{q}\nu}\delta\left(\omega-\omega_{\vec{q}\nu}\right)\sum_{\vec{k}}\left|g_{\vec{k}+\vec{q},\vec{k}}^{\vec{q}\nu,nm}\right|^{2}\label{eq:a2f-def}\\
 &  & \times\delta\left(\epsilon_{\vec{k}+\vec{q},m}-\epsilon_{F}\right)\delta\left(\epsilon_{\vec{k},n}-\epsilon_{F}\right),\nonumber 
\end{eqnarray}
where $g_{\vec{k}+\vec{q},\vec{k}}^{\vec{q}\nu,nm}$ are the matrix
elements of the electron-phonon interaction, $\epsilon_{\vec{k}+\vec{q},m}$
and $\epsilon_{\vec{k},n}$ are the energy of the quasi-particles
in bands $m$ and $n$ with vectors $\vec{k}+\vec{q}$ and $\vec{k}$
respectively. The functional derivative of the critical temperature
 with respect to the Eliashberg function is given by \citep{Bergmann1973}
\begin{eqnarray}
\frac{\delta T_{c}}{\delta\alpha^{2}F\left(\omega\right)} & = & -\left(\frac{\partial\rho}{\partial T}\right)_{T_{c}}^{-1}\frac{\delta\rho}{\delta\alpha^{2}F\left(\omega\right)}.\label{Dev F}
\end{eqnarray}
Now we can calculate the change in the critical temperature and in
the isotope coefficient as \citep{PhysRevB.19.2540} 
\begin{eqnarray}\label{eq 5}
\triangle\ln T_{c} & = & -\int_{0}^{\infty}d\omega\alpha^{2}F\left(\omega\right)R\left(\omega\right)\triangle\ln M,\label{eq:delta-tc-el-ph}
\end{eqnarray}
and
\begin{eqnarray}
\alpha_{el-ph} & = & \int_{0}^{\infty}d\omega\alpha\left(\omega\right).\label{eq:alfa cull}
\end{eqnarray}

The phonon spectrum in PdH(D) is neatly separated in two frequency regions. The Pd vibrations produce acoustic modes while the hydrogen (deuterium) modes give rise to the optical modes. So, by integrating Eq. (\ref{eq:alfa cull}) in the range of frequencies corresponding to Pd we can find the electron-phonon contribution to the isotope effect coefficient corresponding to this atom. In a similar manner we can get the corresponding contribution from hydrogen or deuterium.

Further, the information on the electron-electron contribution can be found from the corresponding Coulomb repulsion parameter $\mu^{*}$. In the Random phase approximation  it is given by \citep{ALLEN19831}
\begin{eqnarray}
\frac{1}{\mu^{*}} & = & \frac{1}{\mu}+\ln\left(\frac{\omega_{el}}{\omega_{ph}}\right).\label{eq:Mu}
\end{eqnarray}
Where $\mu=\left\langle V\right\rangle N\left(E_{F}\right)$ is the product of the average of the coulomb potential and the density of states at the Fermi level, $\omega_{el}$ is an electron energy scale and $\omega_{ph}$ is a phonon energy one. It is therefore evident from Eq. (\ref{eq:Mu}) that $ {\mu^{*}}$ depends on the ion mass through the phonon energy.
 
The isotope coefficient for the electron-electron interaction is given by \citep{LEAVENS19741329} 
\begin{equation}
\alpha_{el-el}=-\frac{d\ln T_{c}}{d\ln M} .
\end{equation}
Here the critical temperature of the isotope is given by \citep{PhysRevB.30.5019}

\begin{equation}
T_{c}^{PdD}=T_{c}^{PdH}+\triangle T_{c}^{el-el} 
 \end{equation}
 where
\begin{eqnarray}
\triangle T_{c}^{el-el} & = & \frac{\partial T_{c}}{\partial\mu^{*}}\left(\mu_{PdD}^{*}-\mu_{PdH}^{*}\right),\label{eq:Delta tc mu}
\end{eqnarray}
and
\begin{eqnarray}
\frac{\partial T_{c}}{\partial\mu^{*}} & = & -\left(\frac{\partial\rho}{\partial T}\right)_{T_{c}}^{-1}\frac{\partial\rho}{\partial\mu^{*}}.
\end{eqnarray}
Now, if we take both contributions into account the total isotope effect coefficient is given by the following equation
\begin{eqnarray}\label{atot}
\alpha_{tot} & = & \alpha_{el-ph}+\alpha_{el-el},
\end{eqnarray}
and the total change in the critical temperature, $T_{c}$, 
\begin{eqnarray}
\triangle T_{c} & = & \triangle T_{c}^{el-ph}+\triangle T_{c}^{el-el}.
\end{eqnarray}
According to Eqs. (\ref{eq:alfa cull} -\ref{atot}), to know $\alpha_{tot}$ it is necessary to know first $\mu^{*}$ and $\delta T_{c}/\delta\alpha^{2}F\left(\omega\right)$. We can get $\mu^{*}$ by solving the linearised Migdal-Eliashberg equation (LMEE) valid at $T_{c}$ ounce $\alpha^{2}F\left(\omega\right)$ is known and then we can calculate the functional derivative, $\delta T_{c}/\delta\alpha^{2}F\left(\omega\right)$, using the formulism of Bergmann \citep{Bergmann1973} and of Leavens \citep{LEAVENS19741329}. For an isotropic superconductor, the LMEE
\begin{eqnarray}
\rho\bar{\triangle}_{n} & = & \pi T\sum_{m}\left[\lambda_{nm}-\mu^{*}-\delta_{nm}\frac{\left|\tilde{\omega}_{n}\right|}{\pi T}\right]\bar{\triangle}_{m}.\label{Eli Ecua}
\end{eqnarray}
Where $\bar{\triangle}_{n}$ is given by 
\begin{eqnarray}
\bar{\triangle}_{n} & = & \frac{\left|\tilde{\omega}_{n}/\omega_{n}\right|\triangle_{n}}{\left|\tilde{\omega}_{n}\right|+\pi T\rho},
\end{eqnarray}
Here $\rho$ is the breaking parameter that becomes zero at $T_{c}$. The
frequency $\tilde{\omega}_{n}$ is 
\begin{eqnarray}
\tilde{\omega}_{n} & = & \omega_{n}+\pi T\sum_{m}\lambda_{nm}sig(\omega_{m}),
\end{eqnarray}
and $i\omega_{n}$ are the Matsubara frequencies, $i\omega_{n}\text{=}i\pi T\left(2n+1\right)$ with $n=0,\pm1,\pm2\ldots$. The coupling parameter $\lambda_{nm}$ is defined as 
\begin{eqnarray}
\lambda_{nm} & = & 2\int_{0}^{\infty}\frac{d\omega\omega\alpha^{2}F\left(\omega\right)}{\omega^{2}+\left(\omega_{n}-\omega_{m}\right)^{2}}.\label{eq:lambdas}
\end{eqnarray}
Notice that $\lambda_{nn}$ is the known electron-phonon interaction parameter.

The Eliashberg function $\alpha^{2}F\left(\omega\right)$ was obtained using the Quantum - Espresso suite code \citep{0953-8984-21-39-395502}. We used the density functional perturbation theory \citep{0953-8984-21-39-395502,RevModPhys.73.515} and the scalar relativistic pseudo potentials of Pardue and Zunger (LDA) \citep{PhysRevB.23.5048}. We used 150 Ry cutoff for the plane-wave basis and a 32 X 32 X 32 mesh for the BZ integrations in the unit cell. For the force constants matrix we used a 16 X 16 X 16 mesh. The sum over $\vec{k}$ in Eq.(\ref{eq:a2f-def}) required a 72 X 72 X 72 grid. To cut the sum over the matsubara frequencies in the LMEE, we used a cutt-off frequency, $\omega_{cutoff} = 10 \omega_{ph}$ where $\omega_{ph}$  is the maximum phonon frequency. 
 
 The electron-phonon  contribution to the change in the critical temperature due to the substitution of hydrogen by deuterium, $\Delta T_c^{e-ph}$, can be computed from Eq.\ref{eq 5}. Now, if we take the experimental critical temperature for PdH to be $T_{c}=8.8\:K$ \citep{PhysRevB.39.4110}, we find that  $\Delta T_c^{e-ph} = -0.332 \:K $ and the contribution due to the electron-electron interaction,  $\Delta T_c^{e-e} = +2.556\:K$. It is obvious that the electron-phonon contribution is almost negligible as compared to the electron-electron one. Furthermore, this theory predicts a critical temperature for PdD 
 \begin{eqnarray}
T_c^{PdD}= T_c^{PdH}+ \Delta T_c^{total}=11.024\:K
\end{eqnarray}
this critical temperature for the isotope PdD is very close to the experimental value of 11.05 K \citep{PhysRevB.39.4110}. Further, we calculated  the isotope effect coefficient contribution to be $+0.0556$ from the electron-phonon interaction and $-0.369$ from the electron-electron interaction. This gives a total isotope coefficient $\alpha^{total}= -0.3134$ which is remarkably near to the experimental value. Further, if we take the experimental values for the critical temperature of PdH and PdD, we obtain an isotope effect coefficient of -0.32889, very close to our findings.

In conclusion, we have shown that if we consider the zincblende crystal structure as the proper one for PdH(D) at low temperatures ($T\ll55$ K), the inverse isotope effect can be explained by taking the electron-electron contribution into account. We found that the electron-phonon contribution is much less important. We can reproduce the experimental measured values for the isotope coefficient and the change in the critical temperature. Our work provides a simple and direct explanation of the observed inverse isotope effect in PdH.

This  work  was  performed  using  the  facilities  of  the  super-computing center (Xiuhcoatl) at CINVESTAV-M\'exico. S. Villa acknowledges the support of Conacyt-M\'exico through a PhD scholarship.

\bibliographystyle{apsrev4-1}
\bibliography{EII_PRL}

\end{document}